\documentclass[conference]{IEEEtran}
\IEEEoverridecommandlockouts
\usepackage{amsmath,amssymb,amsfonts}
\usepackage{algorithmic}
\usepackage{graphicx}
\usepackage{textcomp}
\usepackage{xcolor}
\def\BibTeX{{\rm B\kern-.05em{\sc i\kern-.025em b}\kern-.08em
    T\kern-.1667em\lower.7ex\hbox{E}\kern-.125emX}}
\begin{document}

\title{Gesture recognition with 60GHz 802.11 waveforms}

\author{Eran Hof \and Amichai Sanderovich \and Evyatar Hemo \\ Qualcomm Israel Ltd\\P.O. Box 1212 \\ Israel \\ ehof, amichais, ehemo  @qti.qualcomm.com}

\author{\IEEEauthorblockN{Eran Hof}
\IEEEauthorblockA{\textit{Qualcomm Israel Ltd} \\
P.O. Box 1212 \\ Israel \\ ehof@qti.qualcomm.com}
\and
\IEEEauthorblockN{Amichai Sanderovich}
\IEEEauthorblockA{\textit{Qualcomm Israel Ltd} \\
P.O. Box 1212 \\ Israel \\ amichais@qti.qualcomm.com}   
\and
\IEEEauthorblockN{Evyatar Hemo}
\IEEEauthorblockA{\textit{Qualcomm Israel Ltd} \\
P.O. Box 1212 \\ Israel \\ ehemo@qti.qualcomm.com}}

\maketitle

\begin{abstract}
Gesture recognition application over 802.11 ad/y waveforms is developed. Simultaneous gestures of slider-control and two-finger gesture for switching are detected based on Golay sequences of channel estimation fields of the packets. 
\end{abstract}


\section{Introduction}

In this paper we study gesture recognition techniques applied to a 802.11~ad/y networking chip-set that can be operated with several RF chains simultaneously. The chip-set at hand is operated with two RF chains, one for transmission and one for reception of mmWave communication signals in the band of 60~GHz. These two RF-chains are operated simultaneously to provide radar capabilities. We demonstrate capabilities for slider-control and a switch-operation based on the detection of gentle finger movements. Our study is focused on simple and reliable signal processing techniques that are applied in an open-box fashion with solid engineering motivation. One of the key challenges in providing higher level of user experience is the treatment of certain target instability. These appears as an inherent multiplicative mechanism that is not part of the thermal noise. Our signal processing part provide simple denoising techniques that mitigates this inherent target instability and contribute to excellent user experience. 

\begin{figure}[!t]
\centering
\includegraphics[width=6.0cm]{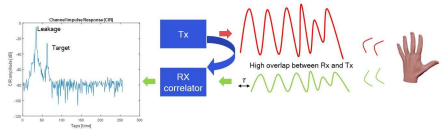}
\caption{A Radar system with a target and the corresponding received signal.} \label{fig:LeakageAndTarget}
\end{figure}

An illustration of the concerned radar system is depicted in Figure~\ref{fig:LeakageAndTarget}. An electromagnetic wave is transmitted from the radar tx module and reflected back from a target object (a hand in Fig.~\ref{fig:LeakageAndTarget}). Some electromagnetic energy is reflected back to the location of the receiver which can sample the received signal and detect the presence of a target. By estimating the time-of-flight and the angle-of-arrival, the radar device can estimate the location and the speed of the the target at hand. 

The paper continues as follows; Section~\ref{sec:taptime} summarizes our tap/time domain notation for operating radar-observation processing of correlation outputs of the 802.11 Golay seqeuences for channel estimation. The studied gesture recognition and related techniques are summarized in Section~\ref{sec:GestureRecognition}.

\section{Notation: On Tap/Time domains} \label{sec:taptime}

\begin{figure}[!t]
\centering
\includegraphics[width=8.0cm]{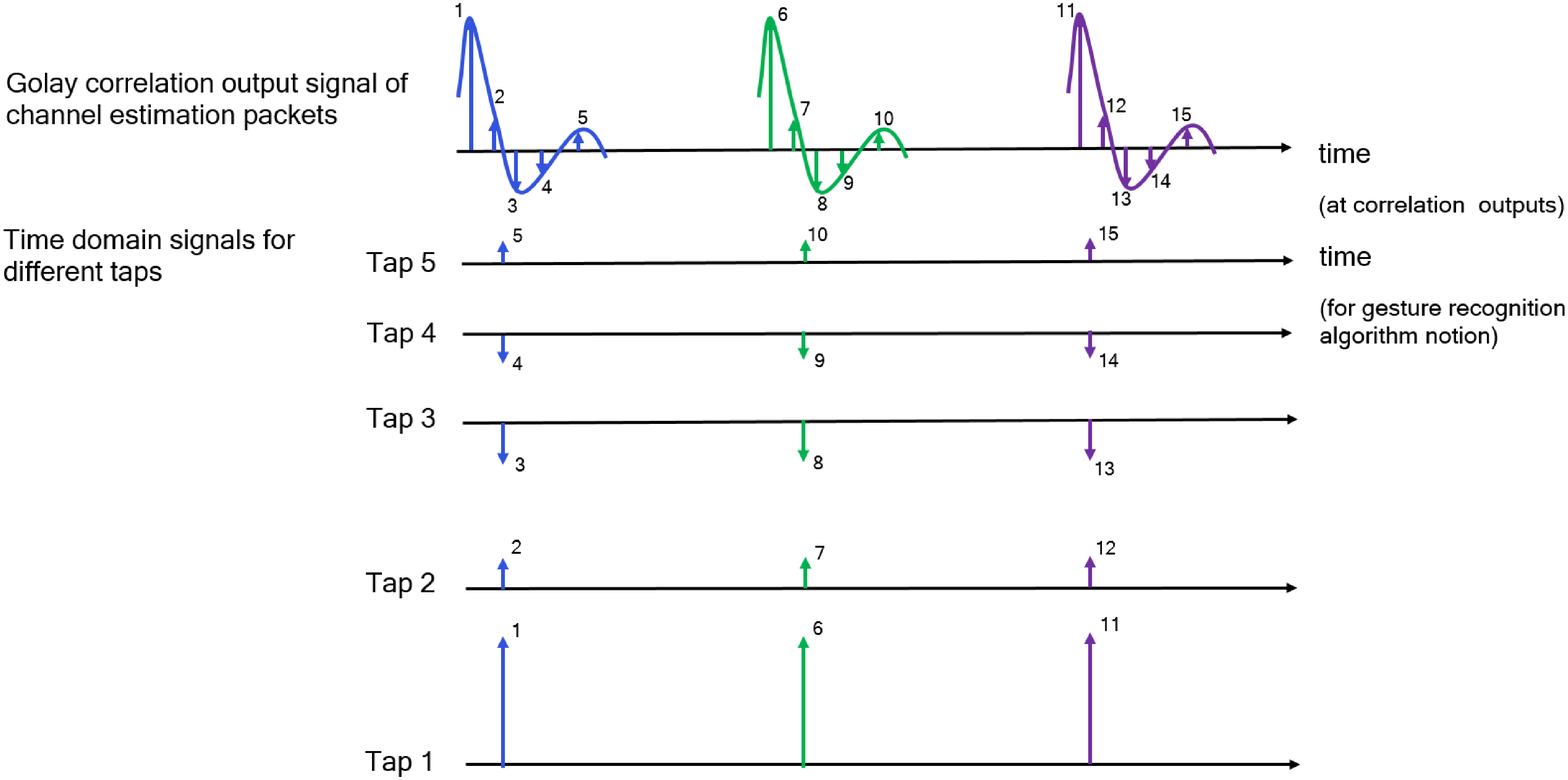}
\caption{Tap and time domains for radar operation of a 802.11 ad/y chip-set. The top-most signal illustrate the Golay correlation output for a duration of 3 pulses. Samples of the correlation output are drawn with arrows labeled 1-15. The tap-notion is illustrated by the 5 time signals, denoted Tap 1 - Tap 5.} \label{fig:tapTimeExample}
\end{figure}

In discussing gesture recognition, we refer to time domain as well as tap domain. Taps are actually time domain samples as well; That is, taps refer to samples of a single packet-correlation of the channel estimation. For the study presented in this paper, channel estimation is carried with complementary Golay sequences as defined by the 802.11 ad/y standard~\cite{ghasempour2017ieee}. In fact, our study is carried with a fully compatible 802.11 ad/y standard compatible networking chip-set that is operated with simultaneous transmitter and receiver RF-chains in order to provide radar capabilities. For the case of radar operation at hand (e.g., gesture recognition, face detection, etc.) instead of a notion of time, the time samples of a single Golay correlation adopt the notion of distance. This is because the time in a perspective of a single correlation corresponds to the wave traveling-time from the setup to the target and back. Hence the reason for the notion of a distance and we therefore refer to the correlation samples as taps. For channel bonding (CB) 1, each tap corresponds to about 8~cm and for CB2 to about 4~cm (at CB1 the bandwidth is 1.76~GHz and for CB2 it is 3.52~Ghz). In our algorithmic study, the notion of a time domain corresponds to complete packets. Our terminology is further illustrated in Figure~\ref{fig:tapTimeExample}. The top-most signal illustrates correlation outputs vs time for the duration of three consecutive Golay pulses. For the purpose of illustration simplicity, each Golay correlation output includes 5 samples. These samples are drawn in Figure~\ref{fig:tapTimeExample} as vertical arrows, labeled as 1-15. In a tap notion, this system produces time signals for 5 taps. These signals are denoted Tap 1 - Tap 5; the time signal for tap 1 is made from samples 1,6 and 11 of the Golay correlation output (topmost illustrated signal), the time signal for tap 2 is made from samples 2, 7 and 12, the time signal for tap 3 is made from samples 3, 8 and 13, the time signal for tap 4 is made from samples 4, 9 and 14 and last the time signal for tap 5 is made from samples 5, 10 and 15. Now, in algorithmic perspective we are dealing with 5 time-signals, each corresponds to a different distance from the setup at hand. Alternatively, we can model our observation signal and a vector signal where in each time, a vector of observations is provided for each tap. That is, in a time $t$, the observed samples is the vector $X_t = (X_t^{(1)},X_t^{(2)},\ldots, X_t^{(N_{\textrm{T}})})$ where $N_{\textrm{T}}$ is the number of taps. That is the number of samples provided for a single Golay correlation. For the illustrative example in Figure~\ref{fig:tapTimeExample}, the sample vector $X_1$ is made from samples 1,2,3,4 and 5, the sample vector $X_2$ from samples 6,7,8,9, and 10 and the third sample vector $X_3$ is made from the samples 11,12,13,14 and 15. As our new time domain corresponds to complete packet, the time sampling interval is the time interval between transmitted packets. 

\section{Gesture Recognition} \label{sec:GestureRecognition}
Slider control requires low latency continuous interface where the user can control the slider level by moving his hand or finger and then see the application slider moves accordingly. We observe the correlation output samples corresponding to distances in the range 0-40~cm from the setup at hand. This range corresponds to the first 5 correlation taps for channel boding (CB)~1and 10 correlation taps for CB~2 (where each tap corresponds to half the distance of a CB1 tap). Gentle finger movements are highly observable in the phase domain of the  received signal\footnote{Note that at 60GHz, the wavelength is 5~mm. Consequently a 1 mm target displacement translates into 144 degrees phase shift (a 1~mm target displacement changes traveled wave distance by 2~mm).}. We therefore collect samples of the phase of the measured Golay correlation outputs (note that the sampled Golay correlation is a series of complex numbers indicative of the traveling waveform distance). 

We set the 802.11ad channel estimation packets to operate at a rate of 2~msec apart. This setting provides samples of the phase of the Golay correlation output in a rate of 500~Hz. To increase SNR, every 2~msec our setup provides 16 consecutive Golay pulses (few micro-seconds apart). These samples are coherently summed up in order to provide a substantial SNR increase per reading without significantly increasing the power consumption. That is, every sample $X = A \cdot e^{j \phi}$, that is provided every 2~msec apart is evaluated according to the following sum:
\begin{equation} \label{eq:snrCombineForPakcets}
X = \sum_{t=1}^{N_{\textrm{p}}} A_t e^{j \phi_t}
\end{equation}
where $N_{\textrm{p}}$ is the number of consecutive channel estimation packets, in our case at hand we set $N_{\textrm{p}}=16$, and $A_t e^{j \phi_t}$ is the $t$-th Goaly correlation output of the $t$-th channel estimation packet out of the $N_{\textrm{p}}$ consecutive micro-seconds-apart packets. For simplicity of the following presentation, we denote (with some abuse of notation) the sample $X$ in~\eqref{eq:snrCombineForPakcets} by $X_t$ to denote it as the $t$-th sample. This time, the notation corresponds to samples taken 2~msec apart (even though it is composed from a sum of faster Golay correlation samples combined for SNR increase). 

In every processing iteration, a processing of a batch of $N_{\textrm{B}}$ samples $X_{t_1},\ldots,X_{t_{N_{\textrm{B}}}}$ takes place, some of these are newly read samples and some are samples from previous iterations. The number of samples in a batch $N_{\textrm{B}}$ and the combination of new and old samples is explained in the following. We are interested mainly in the information carried by the phase; our observations of interest are therefore the $N_{\textrm{B}}$ phase samples $\angle X_{t_1},\ldots,\angle X_{t_{N_{\textrm{B}}}}$. 

\begin{figure}[!t]
\centering
\includegraphics[width=8.0cm]{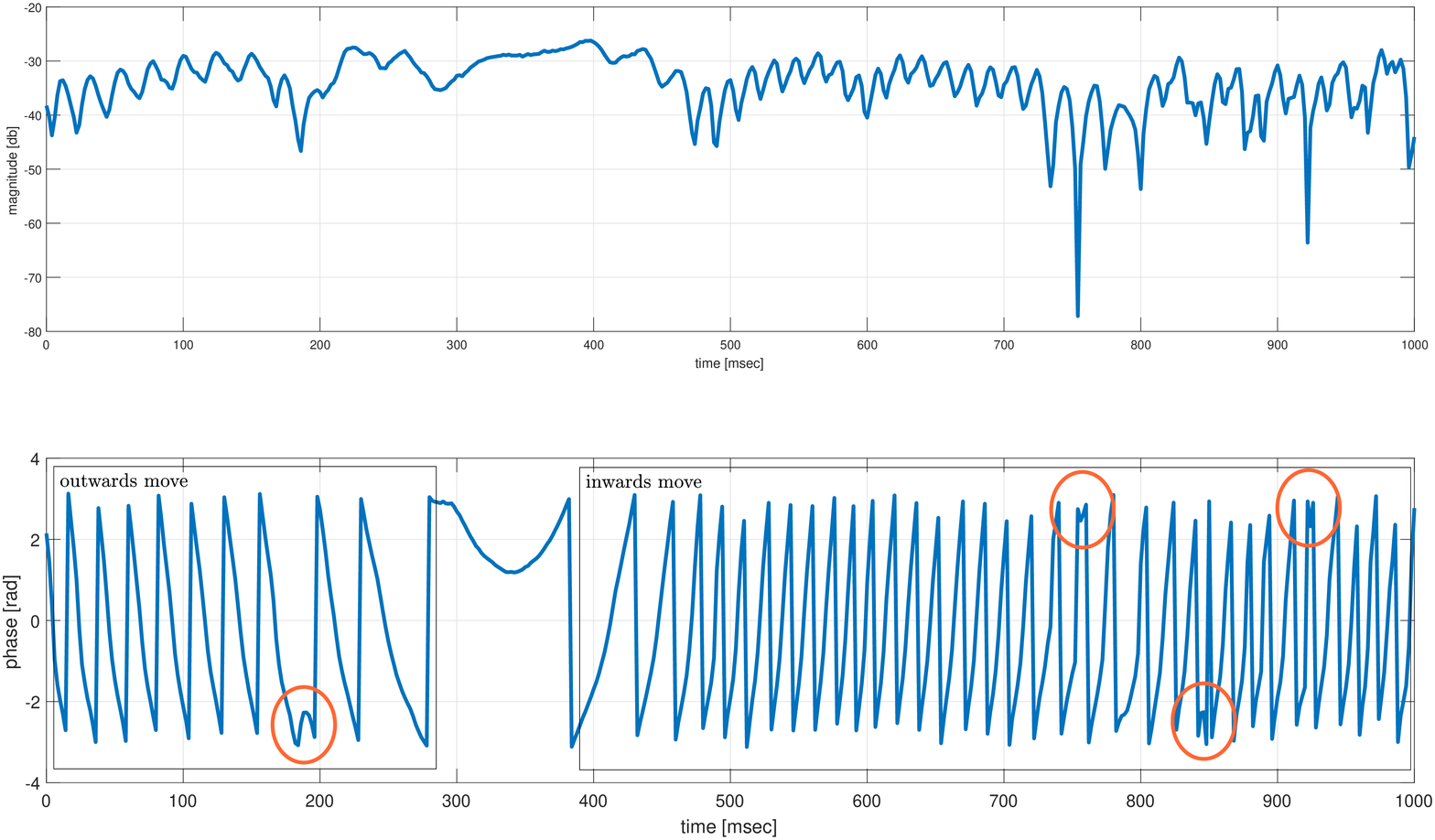}
\caption{One second recordings of a magnitude and phase of a single correlation-tap of a single antenna for a target moving outwards and inwards the setup} \label{fig:singleAntShakExample}
\end{figure}

Our experiments suggest that the target at hand posses movements we are not interested in. These observed movements are not part of the actual gesture and it they  combine to a substantial noisy nature of the sampled observations. This phase noise corresponds to random multiplicative factor of the received signal (in a similar fashion to fadings in communication models). That is, the correlation output, for a given tap, may be modeled as 
\begin{equation*}
X = \alpha A \cdot e^{j \phi} + n
\end{equation*}
where $A>0$ is a real-valued number, corresponds to observation magnitudes (proportional to the RCS), $\phi$ is the phase of the correlation output, $n$ is the thermal noise and $\alpha$ is in general a complex-valued multiplicative noise factor. We further illustrate this noisy mechanism in a sample of a 1 second recoding of a target moving outwards and inwards the setup. We focus on a signal measured from a single correlation tap of a single received antenna. The magnitude and phase of the single-antenna single correlation-tap are shown in Figure~\ref{fig:singleAntShakExample}. We mark the outwards and inwards movement over the phase signal, in general one see a steady slope (either decrease or increase) of the phase. This is the Doppler effect of the movement. In addition, we circled certain events where the phase is shown an abnormal behavior, that may corresponds to a multiplicative noise model that is related to the target instability. 

\begin{figure}[!t]
\centering
\includegraphics[width=6.0cm]{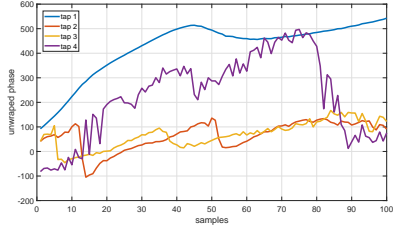}
\caption{Phase samples for Golay correlation output taps} \label{fig:GolayCorrOutput}
\end{figure}

\begin{figure}[!t]
\centering
\includegraphics[width=6.0cm]{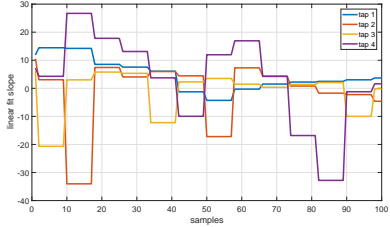}
\caption{Linear fit result for slope of the phase samples of Golay correlation output taps} \label{fig:linearFitSlopesGolayCorrOutput}
\end{figure}

We therefore use two simple but highly effective techniques: first, we apply piece-wise linear fit, and second, we apply a median filtering for the evaluated linear-slopes of the piece-wise linear fitting results. The motivation for piece-wise linear fit follows from the rather steady increase and decrease behavior of the phase that is observed in substantial hand as well as extremely subtle finger movements. Example of such steady increase and decrease are shown in the phase signal in Figure~\ref{fig:singleAntShakExample}. The median filtering is used for denoising the abnormal (multiplicative noise-factor) behavior of the phase. Even an extremely short median window allows to provide steady and clean phase observations for gesture control. We chose to fix the number $N_{\textrm{a}}$ of phase samples required for an evaluation of a single linear fit in the observed sequence. It is found the choice of  $N_{\textrm{a}}=8$ samples per slope enables to provide a fine user-experience. Our study in general with real 802.11ad modules indicates that 5-20 samples at a sample rate of 500-1000 samples per second are fine for the purpose at hand and provides fine user experience in detecting even very delicate slider movements. An example for phase samples and the corresponding slopes is provided in Figures~\ref{fig:GolayCorrOutput} and~\ref{fig:linearFitSlopesGolayCorrOutput}, respectively (the presented plotting are based on real measurements with 802.11ad modules). 

Assume that in a certain iteration we are provided with $N_{\textrm{n}}$ new samples, $X_{t_1},\dots,X_{t_{N_{\textrm{n}}}}$. We therefore choose for this iteration to process a batch of $N_{\textrm{B}}$ samples given by
\begin{equation*}
N_{\textrm{B}} = N_{\textrm{a}} \cdot \left\lceil \frac{N_{\textrm{n}}}{N_{\textrm{a}}} \right\rceil.
\end{equation*}
That is, we take the next integral multiples of $N_{\textrm{a}}$ after $N_{\textrm{n}}$. The missing $N_{\textrm{B}} - N_{\textrm{n}}$ samples are taken from past samples that are logged during the processing. The piece-wise least-square (LS) linear-fit is carried for every $N_{\textrm{a}}$ samples, out of the $\lceil \frac{N_{\textrm{n}}}{N_{\textrm{a}}}\rceil$ groups filled in the current batch of $N_{\textrm{B}}$ samples. For the $N_{\textrm{a}}$ samples $\angle X_{t_1},\ldots,\angle X_{t_{N_{\textrm{a}}}}$, LS linear fit is given by 
\begin{equation*}
X^{\textrm{l}}_{t} = a \cdot t + b
\end{equation*}
where the slope $a$ and constant $b$ are given in general by
\begin{equation*}
a = \frac{\sum_{i=1}^{N_{\textrm{a}}}\bigl(t_i-\bar{t}\bigr)\bigl(\angle X_{t_i} - \overline{\angle X}\bigr)}{\sum _{i=1}^{N_{\textrm{a}}} \bigl(t_i-\bar{t}\bigr)^2}
\end{equation*}
and
\begin{equation*}
b = \overline{\angle X} - a \cdot \bar{t}.
\end{equation*}
Here $\bar{t}$ and $\overline{\angle X}$ are the mean times for the samples at hand and the mean phase, given by:
\begin{equation*}
\bar{t} = \frac{1}{N_{\textrm{a}}} \sum_{i=1}^{N_{\textrm{a}}} t_i
\end{equation*}
and
\begin{equation*}
\overline{\angle X} = \frac{1}{N_{\textrm{a}}} \sum_{i=1}^{N_{\textrm{a}}} \angle X_{t_i}.
\end{equation*}
It is noted that in our problem at hand the linear fit may be further simplified as we are only interested in slopes, not the exact location of the linear fit. Exact location of the linear fit is of no interest in both horizontal and vertical position. As a result the computations of the constant $b$ is redundant and in addition the slopes may be evaluated with respect to arbitrary time units, $t_1=1, t_2=2,...t_{N_{\textrm{a}}}=N_{\textrm{a}}$. We therefore assume that we are provided with a sequence of slopes $\{a_k\}$. These slopes are then filtered by a moving median filter of length $N_{\textrm{m}}$ to provide a sequence of filtered slopes $\{s_k\}$ where the $k$-th filtered sample $s_k$ is given as follows:
\begin{equation} \label{eq:filteredSlopes}
s_k = \textrm{median} \Bigl(a_{k-(N_{\textrm{m}})+1},a_{k-(N_{\textrm{m}})+2},...a_{k}\Bigr).
\end{equation}

For the purpose of operating a slider-control, we tested the following simple tracker 
\begin{equation} \label{eq:simpleTracking}
L_t = \left\{\begin{array}{lr}
		-L  & : {t-1} + \alpha s_t \leq -L \\
		L & : L \leq L_{t-1} + \alpha s_t \\		    
		L_{t-1} + \alpha s_t & : \textrm{otherwise}
  \end{array}
\right. 
\end{equation}
where $L_t$ is the slider-control level at time $t$, $\alpha$ is an attenuation factor, $s_t$ is the current linear fit slope and $[-L,L]$ is the range of operation. Setting the attenuation parameter $\alpha$ was done to maximize the experience among most users. 

It is important to note that the phase samples in a given iteration $X_{t_1},\ldots,X_{t_{N_{\textrm{B}}}}$ are provided by the Golay correlators for all the taps that are observed in a current settings. The taps for which processing should take place is the one corresponding to the actual location of the finger or hand position. However, the exact finger location where the gesture takes place is not known. It is required to provide a mechanism to decide which of the tap samples to use in every iteration. The following three options are suggested:
\begin{enumerate}
\item Take the angles corresponding to maximal strength tap. 
\item Show an average tracker.
\item Update the tracker based on maximal move slope. 
\end{enumerate}
The first option relies on inspecting the Golay-correlation magnitudes $|X_{t_1}|,\ldots,|X_{t_{N_{\textrm{B}}}}|$. Each of these samples is provided for all $N_{\textrm{T}}$ taps. That is,  $X_t = (X_t^{(1)},\ldots,X_t^{(N_{\textrm{T}})})$. We find the tap index $i^*$ that maximizes the magnitude 
\begin{equation*}
i^* = \text{argmax}_{1\leq i \leq  N_{\textrm{T}}} \Bigl( \bigl| X_t^{(i)} \bigr| \Bigr)
\end{equation*}
and evaluate the slopes based on the phase samples corresponding to magnitude-maximizing taps. For the $t$-th sample, we use $\angle(X_{t}^{i^*})$. Taking the strongest tap is by no means optimal in every gesture application. In particular, if gentle finger movements are concerned, it possible that the strongest tap is not reflected from the finger but from the palm of the hand for example (or another large and firmly positioned object). The second option therefore operates on all-taps of interest and shows for the user an average slider position. Variation on this will be to use a quantized slider position where the tracking $L_t$ to initiate the quantization shift is the average on all trackers (with respect to taps). The motivation for this choice is that the tracker that relates to the actual gesture will be the dominant part and therefore the average will resemble the correct tap to follow. Other trackers will either be still, or resembles a noisy random-walk behavior which on average sum-up to zero movement. The third option is to take the tracker that corresponds to the maximal movement. Say we have a tracking value $L_t^{(i)}$, $1 \leq i \leq N_{\textrm{T}}$ for every tap of interest. The second option suggest to use the average tracking $\overline{L_t}$ given by:
\begin{equation*}
\overline{L_t} = \frac{1}{N_{\textrm{T}}} \sum_{i=1}^{N_{\textrm{T}}} L_t^{(i)}.
\end{equation*}
Averaging can be carried even at a slope level; That is compute a single tracker with an average slope. In the tracking equation~\eqref{eq:simpleTracking} we plug-in instead of the slope $s_t$ an average slope $\overline{s_t}$ (where again averaging takes place over the taps). The third option is to use as an update for the tracker as in~\eqref{eq:simpleTracking}  based on the phase $s_t^{(i^*)}$ which corresponds to the maximal movement. That is, let $s_t^{(i)}$, $1 \leq i \leq N_{\textrm{T}}$ be the slopes evaluated for each of the $N_{\textrm{T}}$ taps. We update the tracker based on $s_t^{(i^*)}$ where $i^*$ is given by
\begin{equation*}
i^* = \text{argmax}_{1\leq i \leq  N_{\textrm{T}}} \Bigl( \bigl| s_t^{(i)} \bigr| \Bigr).
\end{equation*}

An important part of the complete user experience relates to the detection of a presence of a finger, its entrance and its leave. Consequently, careful enabling mechanism of the slider-control should be introduced. Reliable detection is provided by testing that slopes do not pass a certain threshold. That is, enable the tracker update in~\eqref{eq:simpleTracking} only when
\begin{equation} \label{eq:enablingSlopesCondition}
| s_t^{(i)} \bigr| < s_{\textrm{Th}} \quad \forall 1 \leq i \leq N_{\textrm{T}}
\end{equation}
where $s_{\textrm{Th}}$ is a predefined threshold to enable the slider. There is some tradeoff in setting $s_{\textrm{Th}}$. If it is made too high, the slider is enabled in cases where it shouldn't be. However, if it is set too low, we may experience cases where a user rapid movements do not trigger slider movement. The detection of a target-presence is easily maintained with observations of magnitude. That is if
\begin{equation*}
\max_{1 \leq i \leq N_{\textrm{T}}} \Bigl( \bigl| X_t^{(i)} \bigr| \Bigr) > M_{\textrm{Th}}
\end{equation*}
then a target presence is decided and the tracking in~\eqref{eq:simpleTracking} takes place. Otherwise, tracking is disabled. Another detection rule which we find to provide an excellent user experience is based on the variation of the magnitudes instead of magnitude. That is, if
\begin{equation*}
\max_{1 \leq i \leq N_{\textrm{T}}} \textrm{std} \Bigl( \bigl| X_{t_1}^{(i)} \bigr|, \bigl| X_{t_2}^{(i)} \bigr|,\ldots,\bigl| X_{t_{N_{\textrm{s}}}}^{(i)} \bigr| \Bigr) > M_{\textrm{Th}}^{\textrm{s}}
\end{equation*}
where std is the standard deviation estimated based on $N_{\textrm{s}}$ recent samples and $M_{\textrm{Th}}^{\textrm{s}}$ is a predefined threshold. It turns out the this detection rule is highly efficient for human hands and fingers due to the natural vibrations of live targets.

\begin{figure}[!t]
\centering
\includegraphics[width=6.0cm]{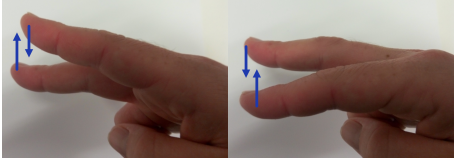}
\caption{Two extreme positions of parallel movement of the index and middle fingers} \label{fig:twoFingerMovement}
\end{figure}

A gesture based on movement of two fingers is considered next. In particular, the detection of switch or increment like commands provided using parallel movement of the index and middle fingers as shown in Figure~\ref{fig:twoFingerMovement} is studied. The two extreme states of this movement are shown in Figure~\ref{fig:twoFingerMovement}; the arrows indicate the direction of movement following the current state. We have applied this gesture detection simultaneous to the slider controller. 

\begin{figure}[!t]
\centering
\includegraphics[width=6.0cm]{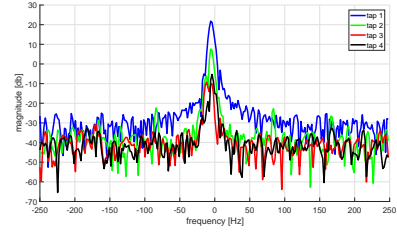}
\caption{Spectrum of Golay correlation with a finger target not in a move} \label{fig:SpectrumSingleFingerNoMove}
\end{figure}

\begin{figure}[!t]
\centering
\includegraphics[width=6.0cm]{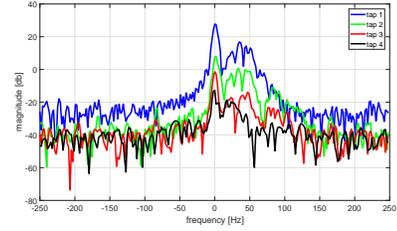}
\caption{Spectrum of Golay correlation with a finger target moving towards in outwards the device (only the index finger is in a move)} \label{fig:SpectrumSingleFingerMove}
\end{figure}

\begin{figure}[!t]
\centering
\includegraphics[width=6.0cm]{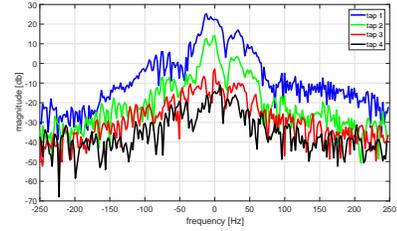}
\caption{Spectrum of Golay correlation with two finger in move (index and middle finger as depicted in fig:twoFingerMovement)} \label{fig:SpectrumTwoFingersMove}
\end{figure}

The detection of the two finger movement is based on spectral analysis. Spectral analysis is carried for a certain tap of interest sampled at the output of the Golay correlation. The motivation for spectral analysis is made clear by examine the signal at output of the Golay correlator, given according to:
\begin{equation*}
e^{j 2\pi \Delta f k}R_{G_{\textrm{a}}}\left(k\right)+e^{j 2\pi \Delta f L + j 2\pi \Delta f k}R_{G_{\textrm{b}}}\left(k\right).
\end{equation*}
Notice that each finger movement introduces a complex exponent with frequency offset $\Delta f$ corresponding to its speed. If Two fingers are moving simultaneously, we should expect to observed this operation in the spectral analysis via prominent energy both in positive and negative frequencies. Let $S_f$ denote the spectrum of the sample at hand at frequency $f$. The following detection rule is applied:
\begin{equation} \label{eq:detectingTwoFingerMov}
\sum_{f \in \mathbb{F}^+} 1_{|S_f|\geq S^{\textrm{th}}} \geq N^{\textrm{th}}_{+} \quad \textrm{AND} \quad \sum_{f \in \mathbb{F}^-} 1_{|S_f|\geq S^{\textrm{th}}} \geq N^{\textrm{th}}_{-} 
\end{equation}
where $\mathbb{F}^+$ and $\mathbb{F}^-$ are the sets of positive and negative frequency bins for spectral analysis, $|S_f|$ is the spectral density at frequency bin $f$, $S^{\textrm{th}}$ is a threshold for prominent energy content at a given spectral bin, and $N^{\textrm{th}}_{+}$ and $N^{\textrm{th}}_{-}$ are threshold for the minimal number of frequency bins, in positive and negative frequencies that are required to be strong enough so that two finger movement is detected. We found that this detection rule is highly reliable. Examples for the spectrum of the Golay correlation with a single target not in a move, a single target in a move and two targets in a move are provided in Figures~\ref{fig:SpectrumSingleFingerNoMove}-\ref{fig:SpectrumTwoFingersMove}. We further suggest that the sets $\mathbb{F}^+$ and $\mathbb{F}^-$ are chosen so that detection is based on frequencies above a certain threshold. That is, 
\begin{equation} \label{eq:positiveFreqs}
\mathbb{F}^+ = \{ f > f^{\textrm{th}}: \  f \in \mathbb{F}^{\textrm{s}}\}
\end{equation}
and
\begin{equation} \label{eq:negativeFreqs}
\mathbb{F}^+ = \{ f < -f^{\textrm{th}}: \  f \in \mathbb{F}^{\textrm{s}}\}
\end{equation}
where $\mathbb{F}^{\textrm{s}}$ is the set of all frequency bins available for spectral analysis\footnote{These are clearly defined by the time-length of the analyzed interval and the sampling rate.} and $f^{\textrm{th}}>0$ is a positive threshold. The reason for this is that a strong prominent energy around DC (zero frequency) is always measured due to the very existence of a strong target (e.g., due to the palm of the hand). Further increase in the reliability of the two finger movement detector in~\eqref{eq:detectingTwoFingerMov} is gained by discarding spectral analysis if highly fast and strong in or out movement is detected based on the signal instantaneous phase. Assume that the current spectral analysis is based on $N_{\textrm{B}}$ samples, $X_{t_1},\ldots, X_{t_{N_{\textrm{B}}}}$, the spectral analysis is discarded if for some $\leq i \leq N_{\textrm{B}}$ it follows that
\begin{equation*}
\angle X_{t_i} > \alpha_{\textrm{th}}
\end{equation*}
where $\alpha_{\textrm{th}}$ is a predefined threshold. Alternatively, one can be provided by inspecting the phase slopes or filtered slopes, e.g. $s_k$ in~\eqref{eq:filteredSlopes} and discard spectral analysis if for some slope evaluated during the spectral interval of interest it follows that
\begin{equation*}
s_k > s_{\textrm{th}}
\end{equation*}
where $s_{\textrm{th}}$ is a predefined threshold for discarding spectral analysis. Another discarding rule is suggested based on the spectral analysis itself. We use the similar spectral detection rule as in~\eqref{eq:detectingTwoFingerMov} with some modifications as follows. First, we apply different frequency threshold $f^{\textrm{th}}$ in~\eqref{eq:positiveFreqs} and~\eqref{eq:negativeFreqs}; in particular, a substantial high frequency is used (in a sharp contrast to modest back-off from DC as applied in two-finger detection). Second, the bin counts $N^{\textrm{th}}_{+}$ and $N^{\textrm{th}}_{-}$ in~\eqref{eq:detectingTwoFingerMov} are set to a substantial higher levels. And last, instead of logical AND, we apply the rule with a logical OR, that is:
\begin{equation*} 
\sum_{f \in \mathbb{F}^+} 1_{|S_f|\geq S^{\textrm{th}}} \geq N^{\textrm{th}}_{+} \quad \textrm{OR} \quad \sum_{f \in \mathbb{F}^-} 1_{|S_f|\geq S^{\textrm{th}}} \geq N^{\textrm{th}}_{-}.
\end{equation*}
As, strong non-related movement are often measured with either strong spectral energy in either positive or negative bands. Last, further increase in detection reliability by looking for consecutive repetition of detection. Specifically, we operate in spectral analysis in moving windows (in time). If several 3-5 consecutive detection rules agree in positive detection for two finger movement, then the detection is set positive.

\section{Conclusions} \label{sec:conclus} Gesture recognition techniques for slide-control and two-finger movement are studied. Our study is based on experimenting with a radar system composed from a chip-set for 802.11 ad/y network communication standard that can operate several RF chains in parallel. The gesture recognition and signal processing techniques are simple, reliable and correspond to solid motivation.

\bibliographystyle{ieeetr}
\bibliography{mybibliography}

\end{document}